\documentclass{article}

\def\myputfigure#1#2#3#4%
{\hskip0.03\textwidth%\hfill
\makebox[0pt]{\hskip#2in
\includegraphics[width=#3\textwidth]{#1}}\vskip#4pt\hfill}

\usepackage{graphicx,emulateapj}

\lefthead{HOLDER ET AL.}
\righthead{INTERFEROMETRIC S-Z GALAXY CLUSTER SURVEY}
\submitted{Published: ApJ, December 1, 2000}

\begin{document}

\title{Expectations For an Interferometric Sunyaev-Zel'dovich Effect Survey
for Galaxy Clusters}

%\shorttitle{An Interferometric SZE Survey}

\author{Gilbert~P.~Holder \altaffilmark{1}, 
Joseph~J.~Mohr \altaffilmark{1,2,3}, 
John~E.~Carlstrom \altaffilmark{1},
August~E.~Evrard \altaffilmark{4} 
Erik~M.~Leitch \altaffilmark{1}}  

%\authoremail{holder@oddjob.uchicago.edu,mohr@rockie.uchicago.edu, 
%   jc@hyde.uchicago.edu, evrard@umich.edu, eml@hyde.uchicago.edu }

\altaffiltext{1}{Department of Astronomy and Astrophysics, 
University of Chicago, 5640 S. Ellis Ave., Chicago, IL 60637}

\altaffiltext{2}{Chandra Fellow}

\altaffiltext{3}{Departments of Astronomy and Physics, University of Illinois,
1002 West Green Street, Urbana, IL 61801}

\altaffiltext{4}{Department of Physics, University of Michigan,
Ann Arbor, MI 48109}

\begin{abstract}
We estimate the expected yield of a
non-targeted survey for galaxy clusters using the
Sunyaev-Zel'dovich effect (SZE).
Estimating survey yields requires a detailed model for
both cluster properties and the survey strategy.
We address this by making mock observations 
of galaxy clusters
in cosmological hydrodynamical simulations.
The mock observatory consists of an interferometric array of 
ten 2.5 m diameter telescopes, operating at a central frequency of 30 GHz 
with a bandwidth of 8 GHz.
For a survey covering 1 ${\rm deg^2}$/month, 
we find that clusters with a mass above 
$2.5 \times 10^{14} h_{50}^{-1} M_\odot$ will be detected at any redshift, with
the exact limit showing a modest redshift dependence. 

Using a Press-Schechter prescription for evolving the number
densities of clusters with redshift, we determine that such a survey 
should find hundreds of galaxy clusters per year,
many at high redshifts and relatively low mass -- an important  regime 
uniquely accessible to SZE surveys. Currently
favored cosmological models predict $\sim 25$ clusters
per square degree. 
\end{abstract}


\section{Introduction}

The evolution of the cluster abundance
is a sensitive probe of the mass density $\Omega_m$
(e.g., \cite{viana99,bahcall98,oukbir97}). X-ray cluster surveys have
started to constrain $\Omega_m$, but they are limited by
their sample size and rapid decline in sensitivity with redshift, 
making the counts very sensitive to the selection function. 
While the selection function
is presumably well-understood, it would be preferable
to have a probe whose sensitivity does not fall off precipitously with redshift.
We will show that a Sunyaev-Zel'dovich effect
survey is ideal in this regard.

Hot ionized cluster gas interacts with passing 
cosmic microwave background (CMB) photons, distorting the
CMB spectrum to create a decrement of CMB flux at lower frequencies and
an excess at higher frequencies. This spectral distortion is independent
of the redshift of the cluster, and only depends on the optical
depth to Compton scattering and the temperature of the gas.
This is the thermal Sunyaev-Zel'dovich effect (SZE) (\cite{sunyaev72}).

There have been several previous predictions of the number of
clusters expected in SZE surveys
(e.g., \cite{bartlett94,barbosa96}), with most earlier work focusing
on the total SZE flux. 
As survey yields depend sensitively on the observing strategy,
we focus here on yields for a proposed interferometric survey 
(see \cite{mohr99a}).

A large catalog of high-redshift clusters that extends to 
low masses would be an extremely useful resource for several reasons. 
Observations suggest that the
universe may no longer be matter-dominated, 
with a large fraction ($\sim 70\%$) of its present energy 
density either in
the form of curvature (open universe) or vacuum 
(cosmological constant) energy.
% Here is where Gus wants structure formation info changed
Linear theory suggests that 
structure formation will slow considerably when the expansion dynamics are no
longer dominated by matter;
this occurred around $z \la 1.5$ if current measurements of $\Omega_m$ are
correct. Therefore, the cluster abundance out to $z \sim 2$ should be
a valuable probe of the matter density of the universe.

A large collection of high-redshift, low-mass
clusters also would provide an ideal sample for exploring evolution of the
intra-cluster medium (ICM) and feedback from galaxy formation. The
shallow potential wells of less massive clusters are 
more strongly affected by energy input from non-gravitational sources
and are therefore the best place to search for the signatures
of such processes.

%In calculating expected yields for an SZE survey, it is necessary to
%determine which properties of a cluster are most important
%for determining its likelihood of detection.

Estimating the cluster yield for an SZE survey requires that we know which
properties of a cluster determine its likelihood of detection.
We will show that the mass of a cluster is the single most important
factor in determining if a given cluster can be detected by
an interferometric survey. In this case, the calculation of the
expected yield separates into two distinct exercises: finding the
minimum observable mass as a function of redshift and calculating
the number density of clusters above a given mass threshold as a
function of redshift.

We determine the minimum observable mass as a function of redshift by
making synthetic observations of N-body+gas hydrodynamical 
simulations (\S\ref{sec:detectability}).
The number density of clusters is calculated using the Press-Schechter
prescription (\cite{press74}; \S\ref{ps-theory}).
Section \ref{sec:results} contains estimates of the expected survey yield
and the results are discussed in Section \ref{sec:discussion}. 

\newpage
\section{Cluster Detectability}
\label{sec:detectability}

The SZE decrement along a given line-of-sight is independent of cosmology.
It is simply proportional to the integrated thermal pressure along the
line-of-sight and therefore only depends on the properties of the
cluster.  The decrement can be written
\begin{equation}
{\Delta T \over T_{\it CMB}} = g(x) \int dl\, n_e(l) {k_B T_{e}(l) \over m_e c^2} 
\sigma_T,
\label{eqn:dt}
\end{equation}
where $n_{e}$ is the electron number density, $T_{e}$ is the electron
temperature,
$m_{e}$ is the electron rest mass, $\sigma_{T}$ is the Thomson 
cross section, $g(x)=x(e^{x}+1)/(e^{x}-1)-4$ with $x=h\nu/k_{B}T_{\it CMB}$ 
and the integral is along the entire line-of-sight;  
by assumption, the only significant contribution to the integral comes from
the cluster atmosphere. In the Rayleigh-Jeans limit ($\nu \ll 200$~GHz) 
the dimensionless frequency factor $g(x)=-2$.

The specific intensity $S_{\nu}$ in the Rayleigh-Jeans regime is
$S_{\nu}=2k_{B}\Delta T\nu^{2}/c^{2} d\Omega$, where $d\Omega$ is the effective
solid angle of the observations.  
Thus the total SZE flux decrement $S_{tot}$ for a galaxy cluster can be written
\begin{equation}
S_{tot}(z) = {2k_{B}^2\nu^2g(x)\sigma_{T}T_{\it CMB}
\over m_{e}c^{4} d_{A}(z)^{2}}\left<T_{e}\right>_{n}
{M_{200}f_{ICM}\over \mu_{e}m_{p}},
\label{eqn:flux}
\end{equation}
where $d_{A}(z)$ is the angular diameter distance,
$\left<T_{e}\right>_{n}$ is the electron density weighted mean 
temperature in the cluster, 
$\mu_{e}$ is the mean molecular weight per electron, $m_{p}$ is the proton
mass, $f_{ICM}$ is 
the ratio of total gas mass to binding mass, 
and $M_{200}$ is a measure of the cluster virial mass,
defined as the mass within $r_{200}$, the radius where
the mean interior density is 200 times the critical density.  
Note that we explicitly ignore contributions to the SZE flux coming from 
outside the virial region.  

Equation 2 indicates that the 
total SZE flux for a cluster is directly 
proportional to the cluster virial mass; 
the only 
dependence on cluster structure is through the density weighted mean
temperature $\left<T_{e}\right>_n$.  
This is unique to a survey with a beam large enough
that the cluster SZE is not resolved. However, for maximum
brightness sensitivity to the SZE effect, the beam used for a
survey should be well matched to the typical angular scale
subtended by clusters. Such a survey will partially resolve 
most of the clusters and will therefore be somewhat sensitive
to the internal cluster structure.
We show below
that for a population of 
clusters with similar ICM mass fraction $f_{ICM}$ and similar 
temperature structure, the detection threshold for an
interferometric SZE survey is also effectively a virial mass limit.

\subsection{Mock Interferometric SZE Observations}

We determine the detection threshold of an SZE survey by analyzing
mock observations of numerical cluster simulations.  These mock observations are
appropriate for a proposed ten-element interferometer, composed
of 2.5~m diameter telescopes outfitted with receivers operating at a central frequency 
of 30~GHz and a bandwidth of 8~GHz.  We assume a correlator 
efficiency of 0.88 and an aperture efficiency of 0.77.  We take the 
system temperature below the atmosphere to be 21~K; this includes 
contributions from spillover past the primary.  We assume an 
atmospheric column with opacity $\tau=0.045$ at zenith, which is a 
conservative estimate appropriate for a low altitude, moderately dry 
site such as the Owens Valley Radio Observatory in summer.  The integration 
time on each cluster is 42~hr, composed of six 7~hr tracks on the 
cluster.  This exposure on each piece of the sky would allow us 
to survey $\sim$1~deg$^{2}$ per month.

Interferometers measure the visibility, $V(u,v)$ which is 
the Fourier transform of the sky brightness distribution multiplied by the
primary beam of the telescopes.
We create mock observations at a 
redshift $z$ by imaging hydrodynamical cluster simulations (described 
in detail below) at that 
redshift; we place those clusters at the appropriate angular diameter distance,
multiply the resulting SZE image  
with the primary beam of the 2.5~m dishes (modeled as a Gaussian with 
${\it FWHM} =14.7'$ at 30~GHz), and then Fourier transform to produce 
visibilities $V(u,v)$.  We then sample these visibilities at the same 
locations in $u$--$v$ space which appear in our simulated 7~hr array track
and add the appropriate noise.

\subsection{Hydrodynamical Cluster Simulations}

The effects of ongoing cluster merging on the ICM density and temperature 
structure can be calculated self-consistently in hydrodynamical simulations. 
Therefore, these simulations provide a way of producing test clusters 
whose complexity approaches that of observed clusters.  
In this work we use an ensemble of 36 hydrodynamical cluster 
simulations carried out within three different cold dark matter
(CDM) dominated cosmologies
(1) SCDM  ($\Omega_m=1$,     $\sigma_8=0.6$, $h_{50}=1$, $\Gamma=0.5$),
(2) OCDM  ($\Omega_m=0.3$, $\sigma_8=1.0$, $h_{50}=1.6$, $\Gamma=0.24$), and
(3) LCDM  ($\Omega_m=0.3$, $\Omega_\Lambda=0.7$, $\sigma_8=1.0$, $h_{50}=1.6$, 
$\Gamma=0.24$).
Here $\sigma_8$ is the power spectrum normalization on $8 h^{-1}$~Mpc scales;
initial conditions are Gaussian random fields consistent with a
CDM transfer function with the specified $\Gamma$ (\cite{davis85}).
These simulations have been used previously for studies of the X-ray emission
from galaxy clusters (\cite{mohr97,mohr99}).

Within each cosmological model, we use two $128^3$
N--body only simulations of cubic regions with scale $\sim400$~Mpc to
determine sites of cluster formation.  Within these initial runs the
virial regions of clusters with Coma--like masses of $10^{15}$~M$_\odot$
contain $\sim$10$^3$ particles.

Using the N--body results for each model, we choose clusters for additional 
study, resimulating them at higher resolution with gas dynamics and 
gravity, as described below.  The size of the resimulated region is set
by the turnaround radius for the enclosed cluster mass at the present epoch. 
The large wavelength
modes of the initial density field are sampled from the initial conditions of
the large scale N--body simulations, and power on smaller scales is sampled
from the appropriate CDM power spectrum.  The simulation scheme is
P3MSPH (\cite{evrard88}), the ICM mass fraction is fixed at 
$f_{ICM}=0.2$, and radiative cooling and heat conduction are ignored.

The high resolution, hydrodynamical simulations of individual clusters 
require two steps:  (1) an initial, $32^3$,
purely N--body simulation to identify which portions of the initial density
field lie within the cluster virial region at the present epoch, and (2) a
final $64^3$, three species, hydrodynamical simulation.
In the final simulation, the portion of the initial density field which ends up
within the cluster virial region by the present epoch is represented using
dark matter and gas particles of equal number (with mass ratio 4:1), while the
portions of the initial density field that do not end up within the cluster
virial region by the present epoch are represented using a third,
collisionless, high mass, species.  The high mass species is
8 times more massive than the dark matter particles
in the central, high resolution region. 
This approach allows us to include the tidal effects of the surrounding
large scale structure and the gas dynamics of the cluster virial region with
simulations that take only a few days of CPU time on a low end UltraSparc.

The scale of the simulated region surrounding each cluster is 
in the range 50--100~Mpc, and varies as $M_{halo}^{1/3}$, 
where $M_{halo}$ is approximately the mass enclosed
within the present epoch turnaround radius.  Thus, the 36 simulated clusters
in our final sample have similar fractional mass resolution; the spatial
resolution varies from 125--250~kpc. We will find that the clusters of
greatest interest to us are those with mass $\ga 2\times 10^{14} M_\odot$,
which have at least several thousand gas particles, even in
the lowest resolution simulations (i.e., the highest mass at $z=0$). 
The masses of the final cluster sample vary by an order of magnitude.
Following procedures described in Evrard (1990), we create  
SZE decrement images along three orthogonal 
lines of sight for each cluster.  Each image is 128$^{2}$ and spans a 
distance of 6.7$h_{50}^{-1}$~Mpc in the 
cluster rest frame.

A strength of using numerical cluster simulations is that 
structural evolution consistent with the cosmological model is  
accounted for naturally by simply examining higher redshift outputs 
of the simulations.

\myputfigure{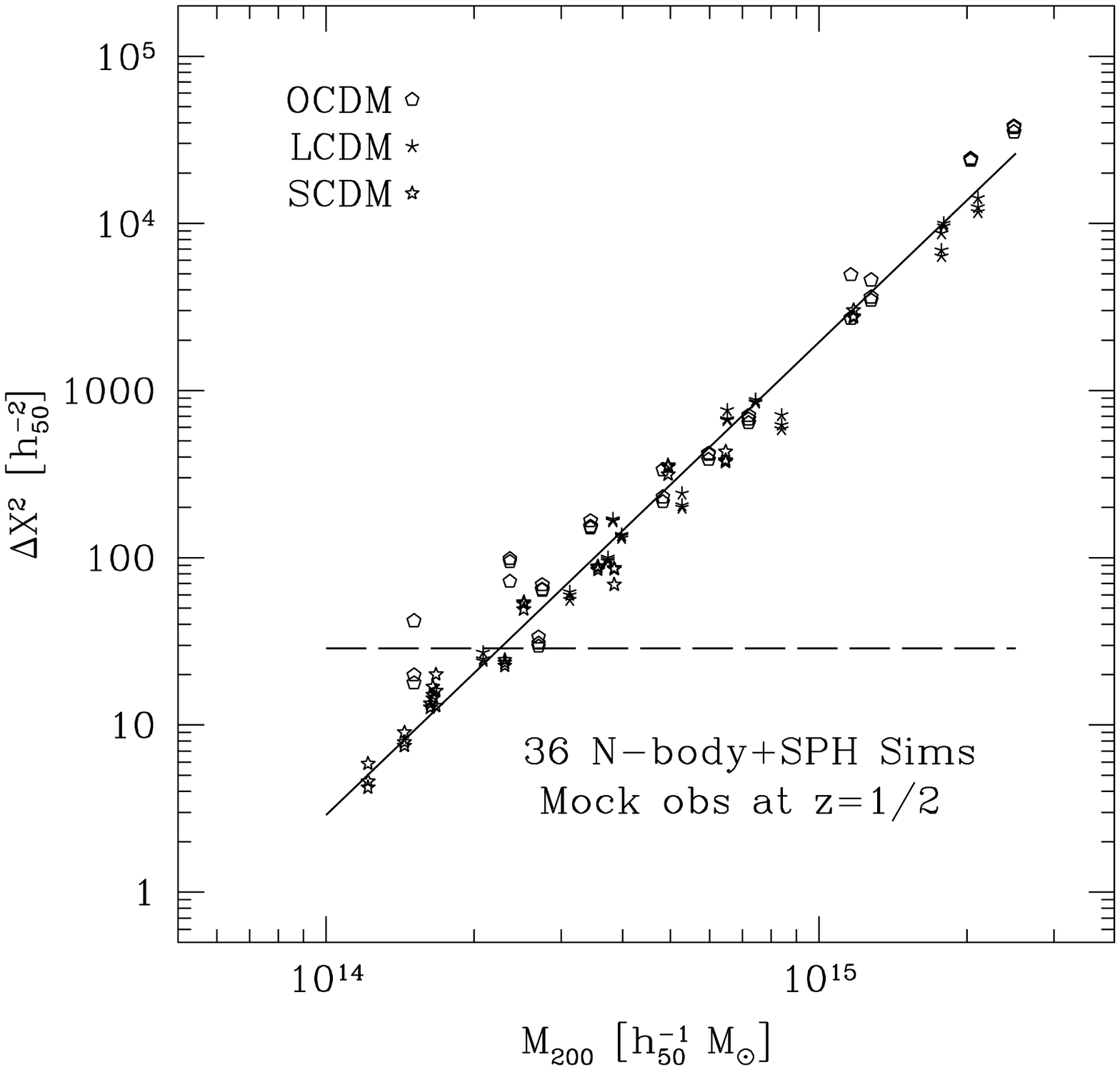}{2.5}{0.5}{-20}
\figcaption{Detection significance, $\Delta \chi ^2$ relative to null model,
as a function
of cluster mass for an interferometric SZE survey. All clusters
were  imaged at a distance corresponding to z=0.5 in an $\Omega_m=0.3$ and
$\Omega_{\Lambda}=0.7$
universe. The dashed horizontal line corresponds to a $5\sigma$ detection for a 
two-parameter fit.
\label{fig:sims-chisq}}

\subsection{Determining the Survey Detection Threshold}

We attempt to detect the clusters in the mock observations by fitting 
the data to the Fourier transforms of spherical $\beta$ models 
(\cite{cavaliere78}), with $\beta=4/3$.  The results are not sensitive
to the value of $\beta$, and in fact a simple Gaussian works well. 
We choose this value of 
$\beta$ because its transform is a simple exponential and because the 
best fits to the SZE decrement images from the simulations yield a mean 
value $\left<\beta\right>=1.1$.  We determine the best fit core radius 
$\theta_{c}$ and central decrement $\Delta T_{o}$ 
by minimizing $\chi^{2}$;  the $\Delta\chi^{2}$ difference 
between the best fit $\beta$ model and the null model (no cluster present) 
provides a measure of the significance of the cluster detection.  
We set our threshold $\Delta\chi^{2}>28.7$ 
(5$\sigma$ for two degrees of freedom). 

Figure \ref{fig:sims-chisq} shows  the detection significance
for 108 mock observations of the
simulated clusters output at redshift $z=0.5$.  Observations of 
clusters in all three cosmologies appear on the same plot, and for 
this example we have imaged all the clusters at the same angular 
diameter distance and accounted for differences in $H_\circ$.  
There is a 
striking correlation between the detection significance and cluster 
virial mass, indicating that even when complex cluster dynamics are 
considered, the survey detection threshold can still be effectively 
described as a mass threshold.  The scatter about this correlation is 
a reflection of the variation in cluster structure due to different 
merger histories and projection effects.  We determine the mass threshold 
by examining the $\Delta\chi^{2}$-$M_{200}$ relation and determining 
the virial mass at which $\Delta\chi^{2}=28.7$.  We calculate the RMS
scatter about the relation at $\Delta\chi^{2}>28.7$, and use that 
scatter as an estimate of the width of the detection threshold in mass;  
the number of mock observations used to characterize the RMS varies
between 21 in our highest redshift bin to over 100 in our low redshift
bins.  We model the scatter as a Gaussian distribution in 
$\chi^{2}$ at each mass.  In this way, the fraction of detected 
clusters at each mass is expressed as an integral over a Gaussian.

Note that the small scatter about the $\Delta\chi^{2}$-$M_{200}$ relation
indicates that mass is indeed the primary factor in determining
cluster detectability.  Moreover, the fact that all three cosmologies 
produce consistent relations once differences in $H_{o}$ are 
accounted for indicates the relative insensitivity of the detection 
threshold to differences in cluster structure.

By repeating this exercise at several redshifts, we determine the 
survey mass threshold as a function of redshift, shown in Figure 
\ref{fig:sims_thresh}.
Note that the thresholds differ for 
each cosmology because of differences in the angular diameter 
distance- redshift relation $d_{A}(z)$.  The error bars on the mass 
limit points indicate the RMS scatter in mass about the 
$\Delta\chi^{2}-M_{200}$ relations (see Fig. \ref{fig:sims-chisq}).  

The hydrodynamical simulations cannot be used to extract mass thresholds 
beyond redshift $z\sim2.3$, because no clusters in our ensemble are 
massive enough to lie above the detection threshold.
For higher redshift and to aid in interpolating the mass threshold
at arbitrary redshift, we use mock observations of
spherical $\beta$ models  
normalized to agree with the simulations (curves in Figure \ref{fig:sims_thresh}).  
We evolve the $\beta$-models in redshift using the spherical collapse 
model (Lahav et al. 1991); this evolution is self-similar, accounting 
for the difference in cluster structure due to evolution of the mean 
cosmological density.  We use these smooth curves essentially as 
fitting functions.

The insensitivity to redshift of the minimum detectable mass in an
interferometric survey follows from a balancing of several effects.  The
increasing angular diameter distance $d_A$ with redshift decreases the
total cluster SZE flux as $d_A^{-2}$ (see equation 2), tending to increase our
limiting mass.  However, this effect is largely offset by cluster
evolution.  At higher redshifts clusters are denser, and, at constant
virial mass $M_{200}$, have higher virial temperatures $T$.  Both of
these effects increase the total SZE flux.  In addition, at higher redshift
a cluster has a smaller apparent size, which enhances the cluster
visibility $V(u,v)$  (Fourier transform of SZE decrement distribution; see
$\S$2.1) at the baselines where we make our measurements.  Taken together,
these effects largely explain the behavior of the limiting mass with
redshift.

\myputfigure{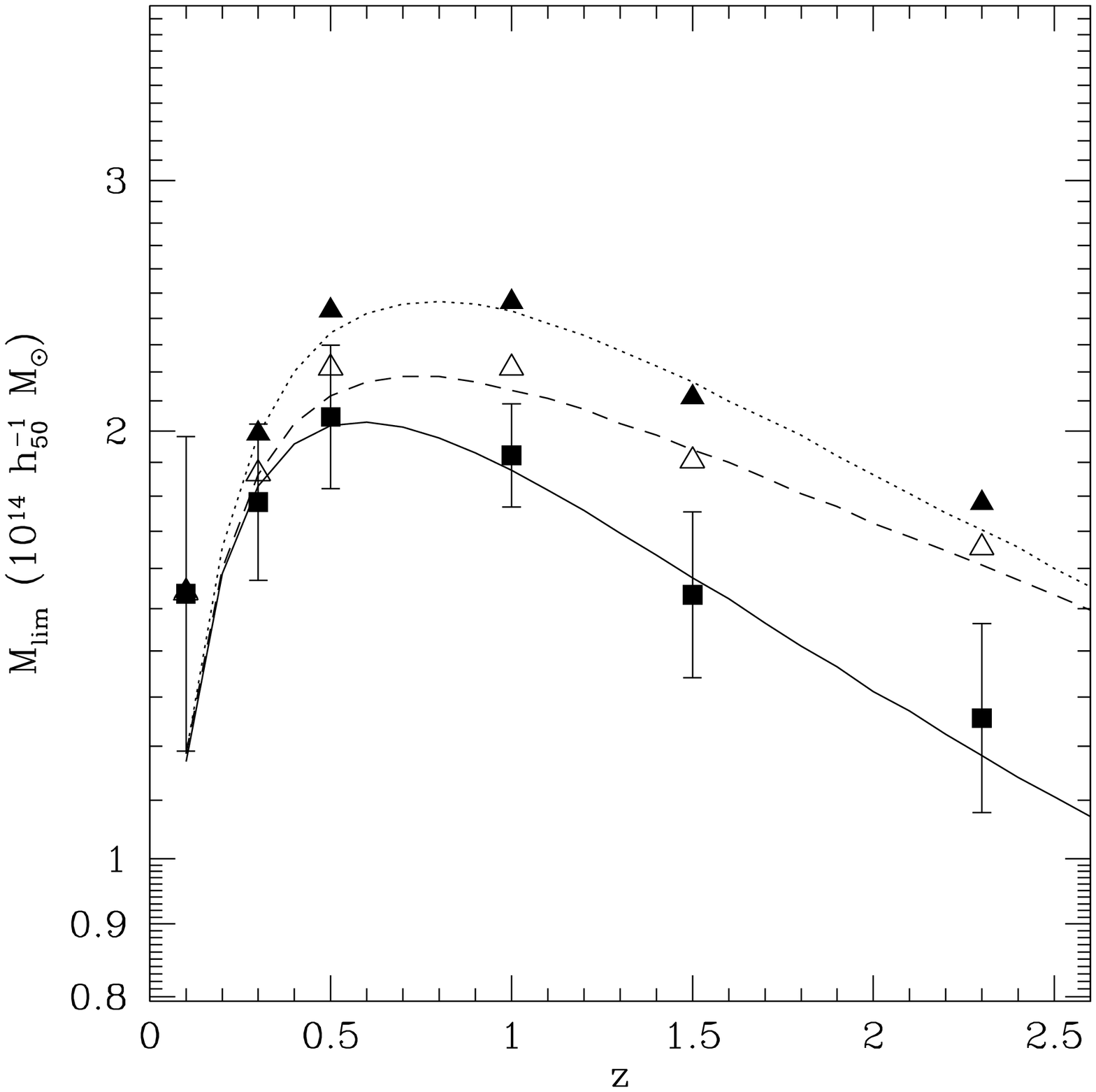}{2.5}{0.45}{-20}
\figcaption{Mass thresholds of detection from the simulations for
our three canonical  cosmologies. 
The differences in mass thresholds between cosmologies  
are simply showing the cosmological 
dependence of the angular diameter distance--redshift relation.
From top to bottom, results are shown for $\Lambda$CDM (solid triangles
and dotted line), oCDM (open triangles and dashed line)
and $\tau$CDM (solid squares and solid line). The uncertainties in the
mass threshold are the same for all cosmologies but, for clarity, are 
only shown for the $\tau$CDM points.
\label{fig:sims_thresh}}

\section{The Cluster Abundance and Its Redshift Evolution}
\label{ps-theory}

Given a minimum
observable mass, one can calculate the number of observed
clusters as:
\begin{eqnarray}
\lefteqn{ N(M > M_{thres}) = } \nonumber \\
& {} c \int d\Omega \int dz \int_{M_{thres}(z)}^\infty 
dM \, {\frac{dn(M,z)}{dM}} {{d_A(z)^2 
(1+z)^2} \over H(z)} \quad ,
\end{eqnarray}
where $d\Omega$ is the solid angle  and $n(M,z)$ is the
comoving number density. 
To calculate the comoving number density of clusters, we used
the Press-Schechter prescription (\cite{press74}).

The comoving number density of bound objects
between masses $M$ and $M+dM$ at redshift $z$ is given by
\begin{equation}
{\frac{dn(M,z)}{dM}} = 
- \sqrt{\frac{2}{\pi}} \frac{\overline{\rho}}{M}\frac{d\sigma(M,z)}{dM}
\frac{\delta_{c}}{\sigma^2(M,z)}
\exp{\left[\frac{-\delta_{c}^2}{2 \sigma^2(M,z)}\right]} \quad , 
\end{equation}
where $\overline{\rho}$ is the mean comoving background density,
 $\sigma(M,z)$ is the variance of the fluctuation spectrum filtered
on mass scale $M$, and  $\delta_{c}$  is the effective
linear overdensity of a perturbation which has collapsed and virialized.
In principle, $\delta_{c}$ has a modest dependence on the cosmological 
density parameter; the spherical collapse model predicts
a variation of only $\sim 5\%$ for $\Omega_m \sim 0.1-1$. 
In this work, we assume a constant
threshold $\delta_c=1.69$ for simplicity (\cite{peebles80}).

Following Viana and Liddle (1999), we take the variance in spheres of radius $R$ to be
\begin{equation}
\sigma(R,z)=\sigma _8(z) \Bigl( { R \over 8 h^{-1} Mpc}\Bigr) ^{-\gamma(R)} ,
\end{equation}
where
\begin{equation}
\gamma(R) = (0.3\Gamma + 0.2)\left[2.92 + \log_{10}
\Bigl({R \over 8 h^{-1} Mpc}\Bigr)\right] \, .
\end{equation}
The comoving radius $R$ is determined as the radius which contains mass $M$
at the current epoch, while $\Gamma$ is the usual CDM shape parameter,
taken to be $0.25$ for this study (\cite{peacock94,dodelson99})
unless stated otherwise; we show
that the results are insensitive to the exact choice of $\Gamma$. 

We examine the cluster abundance in 
three cosmological models:
oCDM ($\Omega_m=0.3, \Omega_\Lambda=0, h=0.65, \Gamma=0.25,
\sigma_8=1.0$), 
$\Lambda$CDM ($\Omega_m = 0.3,\Omega_\Lambda=0.7, h=0.65, 
\Gamma=0.25,\sigma_8=1.0$), and 
$\tau$CDM ($\Omega_m = 1,\Omega_\Lambda=0, h=0.5, \Gamma=0.25,\sigma_8=0.56$), 
with the last model simply a CDM model with the transfer function 
modified to agree with observations of galaxy clustering.
Note that these models differ slightly from the cosmologies assumed for
the simulations.
 
All models are chosen to have a global ICM fraction $f_{ICM}=0.2$, in
rough agreement with observed ICM mass fractions of clusters
(\cite{david95,white95,grego00,mohr99}). 
As can be seen from equation 2, the mass limits will depend on
the ICM mass fraction  and therefore the expected yields will also 
be sensitive to $f_{ICM}$.

We show that the  expected survey yield is very sensitive to 
$\sigma_8$, also calculating the expected yields for oCDM with 
a lower value of $\sigma_8=0.85$ (e.g., \cite{viana99}). 
The constraints on $\sigma_8$ will improve dramatically in the near
future, as new X-ray telescopes are expected to provide a 
much better determination of the local abundance, so we do not expect
uncertainties in $\sigma_8$ to affect interpretation of survey results. 

In a critical density universe ($\Omega_m\!=1,\Omega_{\Lambda}\!=0$),
$\sigma_8 \propto (1+z)^{-1}$.
Following Carroll, Press, and Turner (1992), we 
express growth in alternate cosmologies through a growth suppression
factor which can be approximated as
\begin{equation}
g(\Omega_m,\Omega_{\Lambda}) = {5 \over 2} \ 
{ \Omega_m  \over
\left[ \Omega_m^{4/7} - \Omega_{\Lambda} +
\Bigl(1 + \Omega_m/2 \Bigr)
\Bigl(1 + \Omega_{\Lambda}/70 \Bigr) \right] } \quad .
\end{equation}
In this notation, we can now express the normalization of the power spectrum 
as
\begin{equation}
\sigma_8(z) = {\sigma_8(0)\over 1+z}\,\, {g(\Omega_m(z),\Omega_{\Lambda}(z)) 
\over g(\Omega_m(0),\Omega_{\Lambda}(0))} \quad.
\end{equation}

\section{Expected Survey Yield}
\label{sec:results}

Figure \ref{fig:fuzzy} shows both the differential counts
as a function of redshift and integrated number of clusters
for our three cosmologies.  
Cluster physics is especially uncertain
at high redshift, so we have chosen to cut off all integrals at
a redshift of $z=4$.

\myputfigure{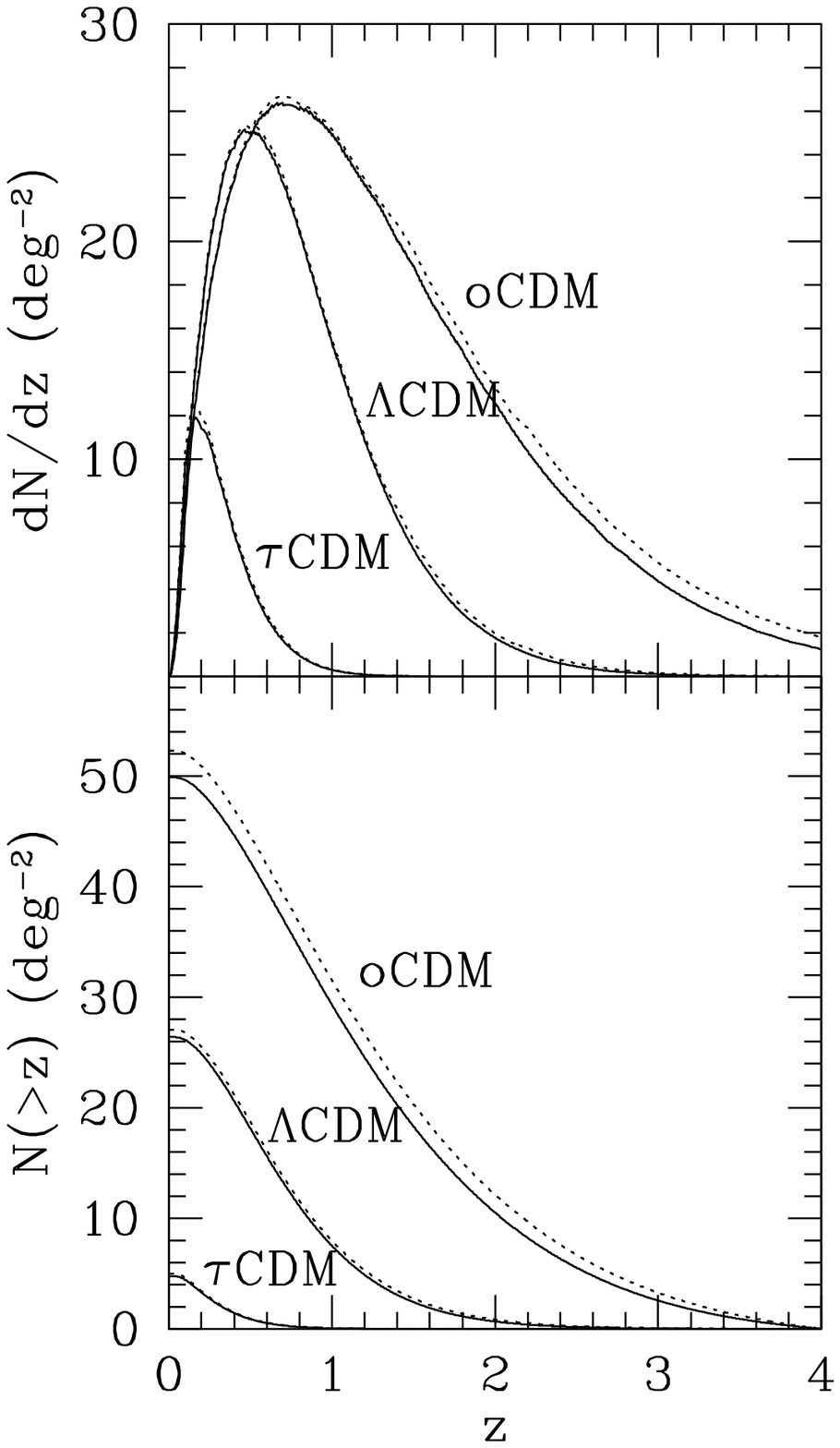}{4.5}{0.65}{-20}
\figcaption[fig3.eps]{Differential counts (top panel) and integrated
counts (bottom) as a function of redshift per square
degree for two different methods of dealing with the limiting mass.
The solid curve for each cosmology assumes a step function
at the best fit limit, while the dotted curve corresponds to a 
Gaussian distribution of $\chi^2$ at each mass.
\label{fig:fuzzy}}
 
These results are quite exciting,
because they indicate that an SZE survey will yield a large
cluster catalog extending to  high redshift.
If the currently favored
$\Lambda$CDM model is correct, then we expect to detect 300 
clusters in a one year survey.  Figure \ref{fig:fuzzy} also shows that the
width of the mass limit has little effect on the cluster
yield.  The yield for each cosmology is shown with two
curves;  the solid curve corresponds to a step function mass
limit where $\Delta\chi^2=28.7$, and the dashed curve
corresponds to a mass limit modeled as a Gaussian
probability distribution in $\Delta\chi^2$ at each mass,
with variance set by the scatter around the
$\Delta\chi^2-M_{200}$ relation seen in the mock
observations.
In particular, the largest differences arise mainly in
the high-redshift region, where there are few clusters
above our mass threshold.
While encouraging, this threshold uncertainty requires 
more attention before future SZE surveys can be correctly interpreted.

\myputfigure{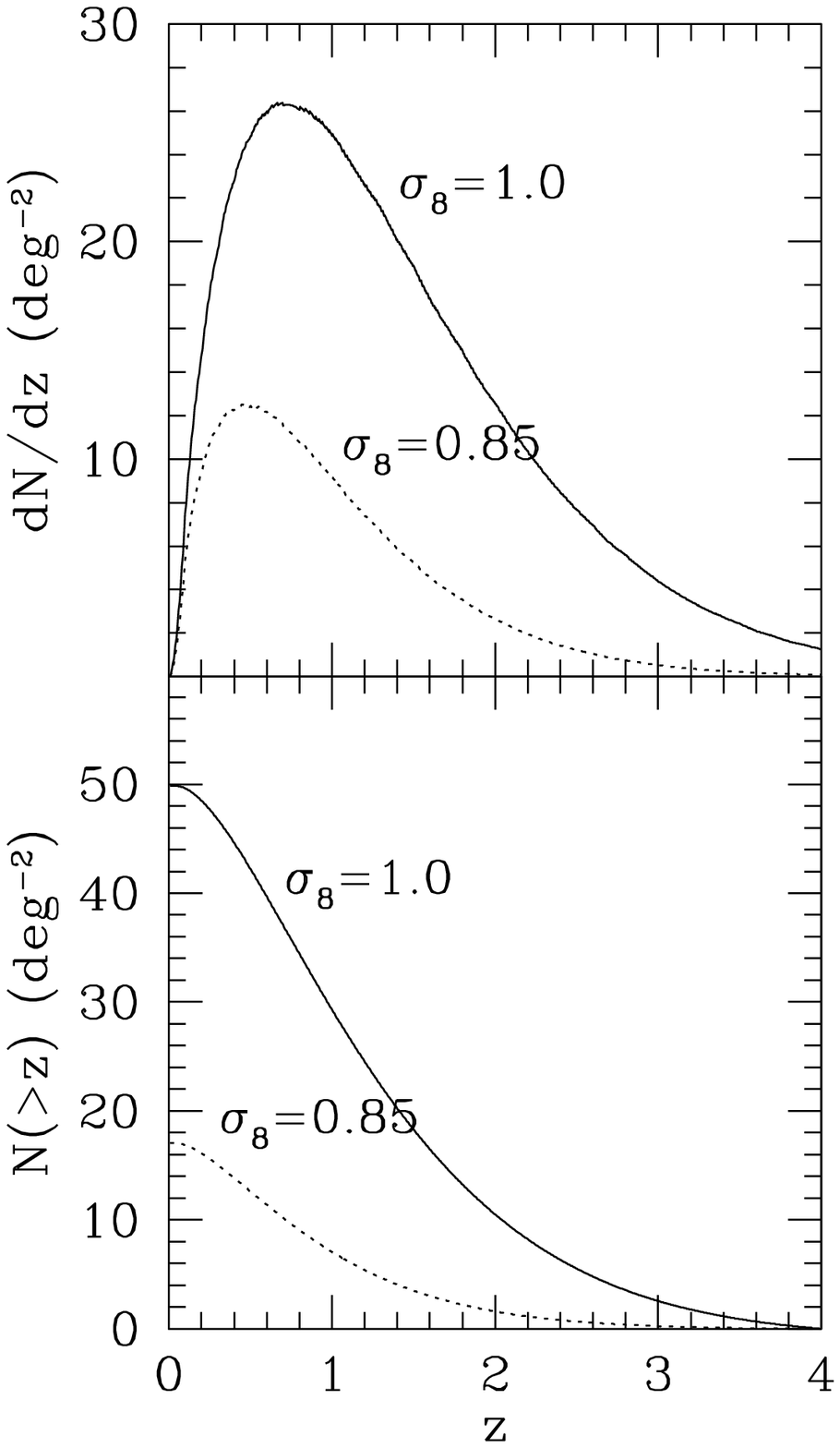}{4.5}{0.65}{-20}
\figcaption[fig4.eps]{Differential (top panel) and integrated (bottom)
expected number of cluster in an oCDM 
model for two values of the normalization of the power spectrum, 
$\sigma_8=0.85$ (dotted) and 1.0 (solid). 
\label{fig:sig_8}}

\vskip 0.1in
The differences between cosmologies at $z \ga 1$ are very promising.
Differences arise because of the different rates of structure growth, 
the dependence of the mass threshold 
on the angular diameter distance, and the differences in the volume element of the
survey. Open models have both a smaller angular
diameter distance and more structure at high redshift, both leading to 
more observed clusters. The effect of
$\Lambda$CDM probing a larger volume is a relatively small effect.

The expected cluster yield depends on cosmology through the
growth rate of perturbations (fixed mainly by the
density parameter $\Omega_m$), the volume per unit solid angle and redshift, 
and the shape and normalization
of the initial power spectrum. 
Varying $\Gamma$, the CDM shape parameter,  
within the $95\%$ confidence interval, $0.2 < \Gamma < 0.3$ 
(\cite{peacock94,dodelson99}),
leads to increases ($\Gamma=0.3$) or decreases ($\Gamma=0.2$)
in the integrated counts of 
$6\%$ for $\Lambda$CDM, $8\%$ for oCDM and $17\%$ for $\tau$CDM. 
The effect is small, indicating that
SZE survey counts alone will not strongly constrain the shape
of the power spectrum.  However, cluster samples constructed through
SZE surveys can be used to study the shape of the power spectrum through
other means, such as the two-point correlation function.

On the other hand, the normalization of the power spectrum,
parameterized by $\sigma_8$, is quite
important for predicted SZE counts, as demonstrated  
in Figure \ref{fig:sig_8}. For clarity, we
have only shown the effect for oCDM.  
The uncertainty in $\sigma_8$ results in a large
uncertainty in the cluster yield.
Upcoming X-ray observations of nearby clusters are expected to 
constrain $\sigma_8$ to
much higher accuracy, and the proposed SZE survey will 
provide an independent measurement. 
Thus, while the $\sigma_8$ uncertainty is significant for predicting 
yields, it will not be a serious impediment to interpreting results from
an SZE survey.
It is important to note that, even for low values of $\sigma_8$, 
we expect to find a significant number of high-redshift clusters.

\section{Discussion}
\label{sec:discussion}

In a low-density universe, half of the clusters
in an SZE survey will have $z\ga 1$. While other
surveys may yield comparable numbers of clusters
at high redshift by surveying a larger fraction of the sky, the SZE
catalog will be unique in having similar sensitivity
at all redshifts; SZE surveys are effectively limited only by the
abundance of clusters above the lowest observable mass. PLANCK 
is expected to find $\sim 10^4$ clusters, but the effective
limiting mass is fairly high, resulting in relatively few high-redshift
clusters (\cite{deluca95,dasilva99}).

An interferometric survey 
with a synthesized beam that partially resolves clusters has
several advantages. 
Point sources are easily identified
and removed from the data; thus, it is unlikely that point sources
will systematically affect the magnitude of the observed
cluster decrement.
In addition, the more massive clusters in our sample
can be imaged with high S/N at the same time as they are detected. 

Extensive follow-up will be
required to make best use of an SZE survey. 
Extensive optical observations will be required 
to obtain redshifts for the detected clusters. This
may be difficult for the highest-redshift objects, but it is
not an insurmountable problem. 
Follow-up with X-ray telescopes would be helpful as well, but the low-mass, 
high-redshift objects are expected to be undetectable in the
X-ray band, even with $10^5 s$ XMM exposures.
However, redshifts, ICM temperatures and X-ray images for a portion of the 
sample would enable direct SZE+X-ray distances, such as those being 
measured with the currently available data (\cite{reese99} and
references therein).  These distances
constrain the angular diameter distance relation,  
which provides an independent measurement of
the cosmological matter density; these results would be 
complementary to those available from consideration
of the cluster counts alone.

% begin CMB discussion
Deep exposures such as the ones considered here could be contaminated 
by primary anisotropies in the CMB.  A minimum separation of 2.5m 
corresponds to multipole $\ell=1571$ at 30 GHz. At this scale, the CMB
anisotropy levels could well be larger than $10 \mu K$, which would
be above the noise levels that we have assumed but well below the
detection threshold.
A shift to higher $\ell$ may be required
to avoid these effects, which can be achieved by 
longer baselines (which could allow larger telescopes)
or a higher observing frequency. We have found that a shift to $\ell\sim3000$
can be easily accommodated with only a small loss in cluster detection 
efficiency.  
The most numerous clusters will be the
low-mass clusters near the detection threshold, which will be the most
compact. Because of this, moving to higher multipoles would not severely
affect our sensitivity to these objects, while this would decrease
possible CMB contamination significantly.
Indeed, larger telescopes may even be slightly more efficient for
detecting high-redshift low-mass clusters, as the smaller primary beam is
better matched to the compact nature of these objects.
% end of CMB discussion  

We use an ICM mass fraction of 20\% in these simulations, which is 
inconsistent with the  
observational constraints,
$f_{ICM} = 0.21 h^{-1.5}_{50}$ (\cite{mohr99,grego00}), 
if $H_\circ$ is significantly higher than $50~{\rm km \, s^{-1}Mpc^{-1}}$. 
However, simple virial 
arguments ($T \propto M^{2/3}$) and Eqn. \ref{eqn:flux} indicate that the
limiting mass scales 
as $f_{ICM}^{0.6}$.  For $H_\circ=80 \, {\rm km \, s^{-1} Mpc^{-1}}$, we expect
$f_{ICM}=0.10$.
While a gas fraction of only 10\% would seriously
reduce the expected number of clusters (by a factor of 2.5-3)
the resulting catalog would still be large and unique.

An uncertainty that we have not
discussed is the effect of galaxy formation 
or preheating on the ICM structure and SZE detectability.
While this is being addressed with numerical simulations 
currently in progress, we do not believe that our results should
depend significantly on gas evolution. ICM evolution mainly
affects the core regions of clusters 
(\cite{ponman99}), on angular scales smaller than those for which the
brightness sensitivity of the interferometer is optimized, 2' to 7'. 
The total SZE flux from the unresolved core depends
only on the temperature and the number of electrons at that temperature
(see \S\ref{sec:detectability}).
Even fairly extreme models of gas evolution lead to only small changes
in the expected counts for a survey sensitive primarily to 
total SZE flux (\cite{holder99b}) and we expect this to be true
for any survey that does not resolve the core regions of clusters.

While the expected cluster yield should be fairly insensitive to ICM evolution,
the properties of the cluster sample and their evolution with redshift
will shed considerable light on the question of evolution of the
ICM. At the same time, such a survey will provide 
determinations of key cosmological parameters that will be entirely
independent of all other determinations.

\vskip 0.2in

\acknowledgements{We are indebted to Erik Reese for his efforts in
developing some of the code that was used for this analysis.
This is work supported by NASA LTSA grant number
NAG5-7986. JEC acknowledges support from the David and Lucile Packard
Foundation and a NSF-YI grant. GPH is supported by 
the DOE at Chicago and Fermilab.  JJM is supported through Chandra Fellowship
grant PF8-1003, awarded through the Chandra Science Center.  The Chandra
Science Center is operated by the Smithsonian Astrophysical Observatory
for NASA under contract NAS8-39073.
AEE acknowledges support from NSF AST-9803199 and NASA NAG5-8458.} 

\newpage

%%%%--References


\begin{thebibliography}{}

\bibitem[Bahcall and Fan 1998]{bahcall98}
Bahcall, N. and Fan, X. 1998, \apj, {504}, 1.

\bibitem[Barbosa {\em et~al.} 1996]{barbosa96}
Barbosa, D., Bartlett, J., Blanchard, A., and Oukbir, J. 1996, \aap, {314},
  13.

\bibitem[{Bartlett} and {Silk} 1994]{bartlett94}
{Bartlett}, J.~G. and {Silk}, J. 1994, \apj, {423}, 12.

\bibitem[Carroll, Press, and Turner 1992]{carroll92}
Carroll, S., Press, W., and Turner, E. 1992, \araa, {30}, 499.

\bibitem[Cavaliere and Fusco-Femiano 1978]{cavaliere78}
Cavaliere, A. and Fusco-Femiano, R. 1978, \aap, {70}, 677.

\bibitem[da~Silva {\em et~al.} 1999]{dasilva99}
da~Silva, A., Barbosa, D., Liddle, A., and Thomas, P. 1999, \mnras, {
  submitted}, (astro--ph/9906289).

\bibitem[David, Jones, and Forman 1995]{david95}
David, L., Jones, C., and Forman, W. 1995, \apj, {445}, 578.

\bibitem[{Davis} {\em et~al.} 1985]{davis85}
{Davis}, M., {Efstathiou}, G., {Frenk}, C.~S., and {White}, S. D.~M. 1985,
  \apj, {292}, 371.

\bibitem[{De Luca}, {Desert}, and {Puget} 1995]{deluca95}
{De Luca}, A., {Desert}, F.~X., and {Puget}, J.~L. 1995, \aap, {300}, 335.

\bibitem[Dodelson and Gaztanaga 1999]{dodelson99}
Dodelson, S. and Gaztanaga, E. 2000, \mnras, {312}, 774.

\bibitem[{Evrard} 1988]{evrard88}
{Evrard}, A.~E. 1988, \mnras, {235}, 911.

\bibitem[Evrard 1990]{evrard90} 
Evrard, A. E. 1990, \apj, {363}, 349.

\bibitem[Grego {\em et~al.} 2000]{grego00}
Grego, L., Carlstrom, J.~E., Reese, E.~D., Holder, G.~P., 
Holzapfel, W~L., Joy, M.~K., Mohr, J.~J., and Patel, S.  2000, 
\apj, {in press}.

\bibitem[Holder and Carlstrom 1999]{holder99b}
Holder, G. and Carlstrom, J. 1999,
\newblock in Microwave Foregrounds, ed. A. de~Oliveira-Costa and M. Tegmark, 
(San Francisco:ASP), 199 .

\bibitem[Lahav {\em et~al.} 1991]{lahav91}
Lahav, O., Rees, M.J., Lilje, P. B., and
Primack, J.R. 1991, \mnras, {251}, 128 .

\bibitem[Mohr and Evrard 1997]{mohr97}
Mohr, J., and Evrard, A.~E. 1997, \apj, {491}, 38.

\bibitem[Mohr {\em et~al.} 1999]{mohr99a}
Mohr, J., Carlstrom, J., Holder, G., Holzapfel, W., Joy, M., Leitch, E., and
  Reese, E. 1999, 
\newblock in {From Extrasolar Planets to Cosmology: the VLT Opening
Symposium}, ed. J. Bergeron and A. Renzini, (Berlin: Springer-Verlag), 150.

\bibitem[Mohr, Mathiesen, and Evrard 1999]{mohr99}
Mohr, J., Mathiesen, B., and Evrard, A.~E. 1999, \apj, {517}, 627.

\bibitem[{Oukbir}, {Bartlett}, and {Blanchard} 1997]{oukbir97}
{Oukbir}, J., {Bartlett}, J.~G., and {Blanchard}, A. 1997, \aap, {320},
  365.

\bibitem[Peacock and Dodds 1994]{peacock94}
Peacock, J. and Dodds, S. 1994, \mnras, {267}, 1020.

\bibitem[Peebles 1980]{peebles80}
Peebles, P. 1980.
\newblock {\em The Large Scale Structure of the Universe}.
\newblock Princeton University Press, Princeton.

\bibitem[{Ponman}, {Cannon}, and {Navarro} 1999]{ponman99}
{Ponman}, T.~J., {Cannon}, D.~B., and {Navarro}, J.~F. 1999, \nat, {397},
  135.

\bibitem[Press and Schechter 1974]{press74}
Press, W. and Schechter, P. 1974, \apj, {187}, 425.

\bibitem[Reese {\em et~al.} 1999]{reese99}
Reese, E.~D. {\em et~al.} 2000, \apj, 533, 38.

\bibitem[Sunyaev and Zel'dovich 1972]{sunyaev72}
Sunyaev, R. and Zel'dovich, Y. 1972, Comments Astrophys. Space Phys., {4},
  173.

\bibitem[Viana and Liddle 1999]{viana99}
Viana, P. and Liddle, A. 1999, \mnras, {303}, 535.

\bibitem[{White} and {Fabian} 1995]{white95}
{White}, D.~A. and {Fabian}, A.~C. 1995, \mnras, {273}, 72.

\end{thebibliography}
\end{document}